\documentclass[superscriptaddress,aps,twocolumn,showpacs,floatfix,amsmath,amssymb]{revtex4-1}

\usepackage{graphics,pstricks,epsfig}

\begin{document}
\newcommand{\be}{\begin{equation}}
\newcommand{\ee}{\end{equation}}
\newcommand{\bq}{\begin{eqnarray}}
\newcommand{\eq}{\end{eqnarray}}
\newcommand{\expect}[1]{\langle #1 \rangle}
\newcommand{\beq}{\begin{equation}}
\newcommand{\beqa}{\begin{eqnarray}}
\newcommand{\eeq}{\end{equation}}
\newcommand{\eeqa}{\end{eqnarray}}
\newcommand{\nbeq}{\begin{equation*}}
\newcommand{\neeq}{\end{equation*}}
\newcommand{\nbeqa}{\begin{eqnarray*}}
\newcommand{\neeqa}{\end{eqnarray*}}
\newcommand{\eps}{\varepsilon}
\newcommand{\fat}[1]{\mbox{\boldmath $ #1 $\unboldmath}}
\newenvironment{eqblock}[2]{\beq\label{#2}\begin{array}{#1}}{\end{array}
                                \eeq}
\newenvironment{neqblock}[1]{\[\begin{array}{#1}}{\end{array}\]}
\newcommand{\beqb}{\begin{eqblock}}
\newcommand{\eeqb}{\end{eqblock}} 
\newcommand{\nbeqb}{\begin{neqblock}}
\newcommand{\neeqb}{\end{neqblock}} 
\def\qsqrt{{\sqrt{2} \kern-1.2em ^4}}
\def\CC{{\rm\kern.24em \vrule width.04em height1.46ex depth-.07ex
\kern-.30em C}}
\def\P{{\rm I\kern-.25em P}}
\def\RR{{\rm
         \vrule width.04em height1.58ex depth-.0ex
         \kern-.04em R}}
\def\id{{\rm 1\kern-.22em l}}
\def\ZZ{{\sf Z\kern-.44em Z}}
\def\NN{{\rm I\kern-.20em N}}
\def\up{\uparrow}
\def\dwn{\downarrow}
\def\L{{\cal L}}
\def\E{{\cal E}}
\def\F{{\cal F}}
\def\ga{{\cal G}[sl(2)]}
\def\H{{\cal H}}
\def\U{{\cal U}}
\def\P{{\cal P}}
\def\M{{\cal M}}
\def\W{{\cal W}}
\def\PACS{\par\leavevmode\hbox {\it PACS:\ }}

\def\a{\alpha}
\def\b{\beta}
\def\d{\dagger}
\def\e{\epsilon}
\def\g{\gamma}
\def\k{\kappa}
\def\l{\lambda}
\def\o{\omega}
\def\t{\tilde{\tau}}
\def\s{S}
\def\D{\Delta}
\def\L{\Lambda}
\def\T{{\cal T}}
\def\TT{{\tilde{\cal T}}}

\def\beq{\begin{equation}}
\def\eeq{\end{equation}}
\def\bea{\begin{eqnarray}}
\def\eea{\end{eqnarray}}
\def\ba{\begin{array}}
\def\ea{\end{array}}
\def\no{\nonumber}
\def\le{\langle}
\def\re{\rangle}
\def\lt{\left}
\def\rt{\right}

\newcommand{\reff}[1]{eq.~(\ref{#1})}

\def\be{\begin{equation}}
\def\ee{\end{equation}}

\def\bea{\begin{eqnarray}}
\def\eea{\end{eqnarray}}

\def\e{\epsilon}
\def\l{\lambda}
\def\d{\delta}
\def\a{\alpha}
\def\b{\beta}
\def\g{\gamma}
\def\s{\sigma}
\def\cb{\bar{c}}

\def\bo{\textbf}
\def\ben{\begin{enumerate}}
\def\een{\end{enumerate}}
\def\cc{^}
\def\bs{\boldsymbol}
\def\ol{\overline}
\newcommand{\vm}[1]{\left< #1 \right>}

\title{Bethe Ansatz approach to the pairing fluctuations in the mesoscopic regime}
\author{Luigi Amico}
\affiliation{CNR-MATIS-IMM $\&$ Dipartimento di Fisica e Astronomia Universit\'a di Catania, C/O ed. 10, viale A. Doria 6
  95125 Catania, Italy} 
 \author{Andreas Osterloh}
\affiliation{Fakult¬at f\"ur Physik, Universit\"at Duisburg-Essen, Campus Duisburg, Lotharstr. 1, 47048 Duisburg, Germany}

\begin{abstract}
We review the exact treatment of the pairing correlation functions 
in the canonical ensemble.
The key for the calculations has been provided by relating the discrete BCS model  
to known integrable theories corresponding to the so called  Gaudin magnets with suitable 
boundary terms. In the present case the correlation functions can be accessed beyond the  formal level, allowing the description of the cross-over
from few electrons to  the thermodynamic limit.
In particular,  we summarize the results on the  finite size scaling behavior of the canonical pairing 
  clarifying some puzzles emerged in the past.
Some  recent developments and applications  are outlined.
\end{abstract}

\maketitle



\section{Introduction}
When a small attractive interaction is switched on in a Fermi gas, 
bound states are  formed non-perturbatively\cite{BCS}.
This is the essence of the BCS pairing phenomenon 
~\cite{TINKHAM,IACHELLO94,RISCHKE,ASTRO},
leading ultimately to a phase transition as a result of a  
competition between the kinetic energy and the tendency of Cooper pairs 
to condense~\cite{BCS,TINKHAM}.  The transition is usually described 
as a  gauge symmetry breaking where the phase of the order parameter 
acquires a fixed value. Nevertheless it should be observed that
because phase and number are conjugate variables, 
a definite value of the phase can be 
consistently reached only in open systems, where the number of Cooper pairs 
can fluctuate. 
{The thermodynamics of macroscopic systems, however, is not 
affected by equilibrium fluctuations} (grand-canonical and canonical statistical
mechanics are equivalent in the thermodynamic limit).  
Accordingly, for macroscopic systems the BCS condensate is a 
quantum coherent state of Cooper pairs characterized by a well defined 
order parameter, which is the Cooper-pair binding energy 
$\Delta$~\cite{BCS,BOGOLIUBOV}.  Away from the thermodynamic limit, 
it becomes delicate when speaking of the existence of the BCS state 
if the BCS pairing energy overlaps only a few energy electronic levels
(the electrons level spacing $\delta \epsilon $ is inversely proportional 
to the volume of the system). The point was famously remarked by 
Anderson with the question: ``What is the size limit for a metallic particle 
to have superconducting properties?''~\cite{ANDERSON}. 
This conceptual challenge has been revived significantly by 
experiments on {\it isolated} metallic grains of nanoscopic size~\cite{BRT}. 
The crucial aspect in the experiments is that the number of electrons 
inside the grain is fixed due to the typically very low capacitance 
of the sample. Therefore the standard superconducting order parameter 
$\langle c^\dagger_\uparrow c^\dagger_\downarrow\rangle$ exactly vanishes
(the mean field approximation in 
the grand canonical ensemble is inappropriate). 
For BCS theory, the quantity playing  the role of the 
``order parameter'' is the  pairing correlation function 
$u_{ij}:=\langle c_{i\dwn} c_{i\up}  c^\dagger_{j\up} c^\dagger_{j\dwn} \rangle$
where $i,j$ are quantum numbers labeling  electronic  energy levels.

Canonical pairing fluctuations were first analyzed 
numerically~\cite{MASTELLONE} in order to study the physics of metallic 
nano-grains\cite{OTHERS}. 
In such mesoscopic regimes the pairing phenomenon appears as a  
cross-over region dominated by superconducting fluctuations 
{sized by the ratio} $\Delta/\delta \epsilon$\cite{MASTELLONE}. 
Because the gauge symmetry cannot be broken at finite sizes,
the system is characterized by these correlations 
(instead of a local observable).
Interestingly, such a physical regime is shared with other important 
physical situations, notably in nuclei~\cite{nuclear}, and more recently 
in  systems of confined degenerate alkali Fermi gases~\cite{coldatoms,bloch_review}.  
Here, we comment that for  $^6 \rm{Li}$ at quantum degeneracy, the mesoscopic 
regime can be achieved  with $\delta \epsilon\sim \Delta\sim 10^{-12} \ \rm{eV}$ 
(corresponding to a confinement frequency of $\omega\sim 1\ \rm{kHz}$)
~\cite{coldatoms,stoof} . In turn, because the BCS order parameter is 
coherent on the length scale $\xi\sim k_F /\Delta$, mesoscopic fluctuations 
in the atomic gas are important for small enough cloud size 
$R\sim \xi$\cite{viverit}. 

Since the system in mesoscopic cross-over regimes is characterized by strong quantum fluctuations, 
its  physical behavior 
is very sensitive to the approximations employed and therefore
 exact results play an important role.
The BCS model was solved exactly in 1964 with the seminal contributions 
by Richardson and Sherman~\cite{RICHARDSON}. The strategy they adopted 
is in the spirit very close to the coordinate Bethe ansatz:  first they  considered  the Cooper pairs as effective bosonic particles; then they were able to   incorporate  the constraint coming from the  actual fermionic statistics into the many-body wave function.  
In 1967  Gaudin~\cite{Gaudin-book} realized that the Richardson solution 
can be obtained with a variation of an approach he had pursued to solve 
the so-called Gaudin magnet~\cite{GAUDIN}.
The Richardson solution remained unnoticed by the condensed matter community
until the late ninety's, when it was re-discovered 
rendering an understanding of the low temperature physics of metallic
grains~\cite{OTHERS} and later on for various applications in 
nuclear physics~\cite{dukelsky_rev} and cold atoms~\cite{bruder}.

Even with the knowledge of the exact solution for the spectrum of 
the system, the computation of the correlation function is a highly 
non-trivial task.
In essence, the complications arise because in the correlation functions 
$\langle \psi |{\cal O}_{loc} |\psi \rangle $ the  eigenstates of 
the Hamiltonian are not easily expressed in terms of the 'natural' states
the operator ${\cal O}_{loc}$ acts on.  
Therefore an exceedingly complicated combinatorial problem arises 
involving (sums of) scalar products between Bethe 
eingenstates~\cite{KOREPIN-BOOK}.
Using the exact eigenstates of the Hamiltonian, the {\it diagonal} 
pairing correlation function $u_{ii}$ was obtained by Richardson 
although it was not evaluated explicitly~\cite{RICHARDSON-CORRELATION}. 
A key progress for accessing the correlation functions exactly came from  
the observation that the BCS model belongs to the class of models  
that can be studied by the powerful techniques developed within the   
Quantum Inverse Scattering Method (QISM)~\cite{KOREPIN-BOOK}.
Specifically, the BCS model is indeed a {\it  twisted}  
Gaudin magnet~\cite{CAMBIAGGIO,AMICO, SIERRA,LINKS} related to 
disordered six vertex models~\cite{SKLYANIN-GAUDIN,AMICO}.  
The first assault on the problem with QISM protocols was done 
in Ref.\cite{amico-osterloh} where various correlation 
functions have been computed. 
A major simplification of the formulas was achieved by 
Zhou {\rm et al.} ~\cite{LINKS} by application of the so called 
Slavnov formula for the calculation of the scalar products between 
Bethe states~\cite{slavnov} together with the 
'solution of the inverse problem' \cite{kmt98,mtgk}. 
Correlation functions have been expressed as the sum of certain determinants. 
The ultimate progress has been achieved by Faribault, Calabrese and Caux  
who further reduced the complexity involved in the calculations, by 
applying certain reduction formulas for the determinants\cite{calabrese_static}. 

The aim of this article is to  review the path we summarized above.
In Section II we intoduce the BCS model and highlight its integrability.
In Section III, we comment on the derivation of correlation functions by means
of generating functions. Section IV is devoted to determinant representations
of correlation functions making use of the solution to the inverse problem together with the reduction formulas worked out in \cite{calabrese_static} . 
Section V is focused on the pairing amplitute in the canonical ensemble.  Section VI has
a short view on the thermodynamic limit of the model. Section VII presents further ramifications.




\section{The BCS model}
\label{model}
The BCS Hamiltonian is 
\begin{eqnarray}
H  = \sum_{j=1 \atop \sigma= \up, \dwn}^\Omega 
\varepsilon_{j} n_{j \sigma}
  -g  \sum_{i, j=1}^\Omega   c_{i\up }^\dagger c_{i \dwn}^\dagger 
c^{}_{j \dwn} c^{}_{j \up} .
\label{pairing}
\end{eqnarray}
$g$ is the pairing coupling constant; 
the quantum numbers $j\in \{1\dots \Omega\}$  label the 
single particle energy levels
$\epsilon_j$ which are doubly degenerate since  
$\sigma \in \{\up, \dwn\}$ labels 
electron spin states; 
$c_{j,\sigma}$ and $n_{j,\sigma}:=c_{j \sigma}^\dagger c_{j \sigma}$ 
are annihilation and number operators, respectively. 

In the following we explain the connection of the BCS model with the 
$sl(2)$-Gaudin model. For this goal we 
introduce the fundamental  realization of   
$su(2)\simeq sl(2)$ in terms of electron pairs 
$
S^-_j := c_{j,\dwn} c_{j,\up}, \;\; S^+_j:=(S^-_j)^\dagger= 
c^\dagger_{j,\up} c^\dagger_{j,\dwn} , \;\; 
S^z_j := (c^\dagger_{j,\up} c_{j,\up}+ c^\dagger_{j,\dwn} c_{j,\dwn}-1)/2 
$. 
The  $sl(2)$  ``lowest''  weight module  is generated by the vacuum
vector $|0\rangle_j $, 
$ 
S^-_j |0\rangle_j=0\;, \quad S^z_j |0\rangle_j=s_j |0\rangle_j
$,
where $s_j$ is the ``lowest''  weight ($s_j=-1/2$ for spin $1/2$, 
which is the case of interest here~\cite{SKLYANIN-NOTATION}). 
The quadratic Casimir operator is
$
S_j^2:= (S^z_j)^2+
{{1}\over{2}}\left(S^+_j S^-_j + S^-_j S^+_j\right ) 
$, 
$S_j^2|0\rangle_j=s_j (s_j-1) |0\rangle_j$.
The bilinear combinations $S^+_j S^-_j$ and  
$S^-_j S^+_j$ can be expressed in terms of Casimir and Cartan 
operators 
\begin{equation}
S^\pm_j S^\mp_j=S^2_j-(S^z_j)^2 \pm S^z_j\;. 
\label{pairing-operators}
\end{equation}
The key observation is that the constants of the motion of the model 
Hamiltonian (\ref{pairing}) can be obtained by the QISM. 
The method follows  an 'inverse' procedure to obtain a Hamiltonian that is integrable by construction. The starting point 
is to find a couple of matrices, $L$ and $R$ satisfying the Yang-Baxter equation
\begin{equation}
R(u-v) L(u)\otimes L(v)=L(v)\otimes L(u) 
R(u-v)\;,
\label{RL}
\end{equation}
where $u\in C$ is the spectral parameter.
For the present case, the relevant matrices can be obtained through 
\begin{equation}
R_{X}(u-z;\eta)=\id \otimes \id + 
f(u-z,\eta)
\fat{\sigma} \otimes \fat{X}  \;, 
\label{R-matrix}
\end{equation}
where $\fat{\sigma}$ is the vector of Pauli matrices, and  
$f(x,\eta):= 2\eta/(\eta -2 x)$ depending on the arbitrary parameter 
$\eta \in \RR$.
The $R$--matrix corresponds to $\fat{X}=\fat{\sigma}$ and $z=0$  
in (\ref{R-matrix}) while  the Lax matrix $L$ is obtained as $R_{\fat{S}}$
\begin{equation}
L_j(u):= \left(
\begin{array}{lcc}
\id+f(u,\eta) S^z_j &  f(u,\eta) S^-_j  \\
 f(u,\eta) S^+_j  & \id  -f(u,\eta) S^z_j 
\end{array}
\right)\, .
\end{equation}

$R_X$ defines the column-to-column and row-to-row scattering matrices, 
respectively, of the two dimensional six vertex model  with inhomogeneities 
${\mathbf \epsilon}=\{\epsilon_1,\dots,\epsilon_\Omega\} $~\cite{KOREPIN-BOOK}.  
The monodromy matrix 
\beq
T (u|{\bf \epsilon}) := L_\Omega (u-\epsilon_\Omega)\dots
 L_1(u-\epsilon_1)
\eeq
satisfies the Yang-Baxter equation
\beq
R(u-v) T(u|{\bf \epsilon} )\otimes T(v|{\bf \epsilon})=T(v|{\bf \epsilon})\otimes T(u|{\bf \epsilon}) 
R(u-v)\;.
\eeq   
The {\it twisted}  monodromy matrix 
\beq
\tilde{T}(u| {\bf \epsilon}):= 
\left (\exp{a(\eta)\sum_j { \sigma^z_{j}}}\right ) T(u| {\bf \epsilon})
\eeq
is then
\begin{equation}
\tilde{T}(u| {\bf \epsilon})=
\left ( \begin{array}{cl} 
A(u|{\bf \epsilon}) & B(u|{\bf \epsilon})\\
C(u|{\bf \epsilon}) &  D(u|{\bf \epsilon}) 
\end{array} \right ) \; .
\end{equation}
It satisfies the Yang-Baxter equation as well due to
$\left [e^{a(\eta) \sigma^z_{j}}\otimes e^{a(\eta) \sigma^z_{j}}\, , \; R\right ]=0$.
The transfer matrix is the trace (over the auxiliary $2\times 2$ space) of 
the monodromy matrix  
\begin{equation}
\tilde{t} (u|{\bf \epsilon}):=tr_{(0)}\tilde{T}(u| {\bf \epsilon})=A(u|{\bf \epsilon}) + D(u|{\bf \epsilon}) \;. 
\label{monodromy}
\end{equation}
The latter is a generating function of integrals of the motion of 
the theory because they commute at different values of spectral parameters:
$[\tilde{t} (u|{\bf \epsilon}),\tilde{t} (v|{\bf \epsilon})]=0$.
 The expansion $\tilde{t} (u|{\bf \epsilon})=\sum_{a=0}^\infty \eta^a \tilde{t}_a 
(u|{\bf \epsilon})$
generates a hierarchy of integrable systems since
\begin{equation}
\sum_{a=b+c=0}^\infty [\tilde{t}_b (u|{\bf \epsilon}),\tilde{t}_c (v|{\bf \epsilon})]=0 
\; .
\label{hierarchy}
\end{equation}
{ The sum is on ordered partitions of $a$ including $b\,  \vee\, c=0$.}

The first non trivial terms of the transfer matrix $\tilde{t}(u|\eps)$ are
\begin{equation}
\tilde{t}(u|{\bf \epsilon})=2 \id +2 \eta^2 \sum_{j=1}^{\Omega} {{\tau_j}\over{u-\epsilon_j}} \; ,
\label{transf}
\end{equation}
where
\begin{equation}
\label{twisted_gaudin}
\tau_l = S^z_l/g -  \; \Xi_l\;, \quad \Xi_j:= \sum_{l\neq j}^\Omega 
{\frac{ {\fat{S }}_j \cdot {\fat{S}}_l }{( \varepsilon_j- \varepsilon_l)}}  
\end{equation}
and ${\fat S}_j:= (S^x_j, S^y_j,S^z_j)$  are spin vectors; 
$S^\pm_j=1/\sqrt{2}(S^x_j\pm i S^y_j)$).
The operators (\ref{twisted_gaudin}) define the twisted Gaudin magnet, where $[\tau_j,\tau_l]=0$ holds.
By these integrals of the motion, the 
BCS model becomes connected with the Gaudin Hamiltonians 
$
\Xi_j$ as the  Hamiltonian~(\ref{pairing}) can be 
expressed in terms of $\tau_j$ as
\begin{equation}
\label{tau-H}
 H = g \sum_{j=1}^\Omega 2 \varepsilon_j \tau_j +  g^3\sum_{j,l=1}^\Omega \tau_j  \tau_l  + const.
\end{equation}
which is manifestly integrable~\cite{CAMBIAGGIO}:
\begin{equation}
[H, \tau_j] = 0\;,  j =1, \dots, \Omega\; . 
\end{equation}

The exact eigenstates of the BCS  model~\cite{RICHARDSON,GAUDIN} 
are obtained  first by diagonalizing $\tilde{t}(u)=A(u|\epsilon)+D(u|\epsilon)$:
\begin{equation}
\label{spectral}
\tilde{t}(u|{\bf \epsilon}) |\psi ( \{ u_i \} |{\bf \epsilon})\rangle =
\Lambda ( u, \{u_i \}|{\bf \epsilon})
|\psi (  \{u_i \}|{\bf \epsilon}) \rangle
\end{equation}
 where the Bethe vectors  are 
 \begin{equation}
 |\psi(\{u_i\})\rangle 
 =\prod_{\alpha=1}^N B(u_\alpha)|0\rangle \; .
 \end{equation} 
Then the eigenstates of  $\tau_j$\cite{SKLYANIN-GAUDIN} are obtained 
by the quasiclassical expansion of the Eq.~(\ref{spectral}) with 
$|\psi\rangle =|\psi_0\rangle +\eta |\psi_1\rangle $.  $|\psi_1\rangle $ results to be 
\begin{eqnarray} 
\label{BA-state} 
|\psi_1\rangle  = \prod_{\alpha = 1}^N &&S^+(e_\alpha) |{\rm 0} \rangle \; ,
\end{eqnarray}
where 
\beqa
S^\pm(u) &:=& \sum_{j=1}^\Omega \frac{S^\pm_j}{ u-2 \varepsilon_j } 
\; , \; 
S^z(u) := \sum_{j=1}^\Omega \frac{S^z_j}{u-2 \varepsilon_j} \; ,
\label{gaudin-operators}\\
s(u)&:=&\sum_{j=1}^\Omega {s_j}/{(u-2 \varepsilon_j)} \; .
\eeqa
and the rapidities $u_\alpha= u_0+\eta e_{\alpha}$ with $e_\alpha$   given by the Richardson equations
\begin{equation}
\label{re}
s(e_\alpha) =\frac{1}{2g } + \sum_{\beta=1 \atop \beta\neq \alpha}^{N} \frac{1}{ e_\beta- e_\alpha} 
\; , \quad \alpha = 1, \dots , N  \; .
\end{equation}
Finally the eigenvalues of $H$ are obtained via (\ref{tau-H}):
$ H| \E\rangle_N 
= E |\E \rangle_N  
$ with $\; E = \sum_{\alpha =   1}^{N} e_\alpha$.
Throughout the article we will consider the half filling case 
$N=\Omega/2$, unless it is stated differently.

We observe that the operators (\ref{gaudin-operators})
span the infinite dimensional Gaudin algebra 
${\cal G}[sl(2)]$. 
The lowest weight module of ${\cal G}[sl(2)]$ is generated by the vacuum 
$|0\rangle:= \otimes_{j=1}^\Omega |0\rangle_{j}$:
$
S^-(u) |0\rangle=0 
\; , \quad S^z(u) |0\rangle
=s(u) |0\rangle \; ,
$
wheren $s(u)$ 
is the lowest  weight  of ${\cal G}[sl(2)]$.
We observe that the  integrability of the BCS model can be obtained as an algebraic property of  ${\cal G}[sl(2)]$.
In fact the mutual commutativity of $\tau_j$ descends from the 
relation between 
$\tau(u):= \sum_{j=1}^\Omega {{\tau_j}/{(u-2 \varepsilon_j)}}$ ($\tau_j$ are residues of 
	$\tau(u)$ in $u=2 \varepsilon_l$) and 
invariants (trace and quantum determinant\cite{SKLYANIN-GAUDIN}) 
of $\ga$:
\begin{equation}
\tau (u) =c(u) + s^{[2]}(u) 
\end{equation}
where $c(u)$ is a twisted Casimir operator 
 \begin{eqnarray}
c(u):= &&\frac{1}{g}S^z(u)+ \\ 
&&2\left [ S^z(u)S^z(u) + {{1}\over{2}}\left( S^+(u)S^-(u)+S^-(u)S^+(u) \right ) \right]  \nonumber 
\end{eqnarray}
and
\beq
s^{[2]}(u):=\sum_{j=1}^\Omega {{s_j }/{(u-2 \varepsilon_j)^2}}
\; .
\eeq
The  property  $[c(u),c(v)]=0$ is the origin of the integrability
of the BCS model. Therefore finding the spectrum of the BCS model means finding the representations of a twisted Gaudin algebra (labeled by its  Casimir operator).

We mention that the Richardson equations~(\ref{re}) are intimately related to  
the algebraic structure of ${\cal G}[sl(2)]$ 
in that they act as constraints on the lowest weight $s(e_\alpha)$. 
Thus,  the difference between the BCS and Gaudin model 
amounts to a different constraint imposed on the lowest weight 
vector of ${\cal G}[sl(2)]$ which leads to different sets ${\cal E}$, 
${\cal E}'$ ($\E' $ is spanned by the solutions of ~(\ref{re}) 
when $g\to \infty$).
We will use this fact to extend the Sklyanin theorem 
~\cite{SKLYANIN-CORRELATORS}
to the BCS model.  

In the next sections we will be focusing on the following $M$-point charge and pairing correlation functions (CF) 
\vspace*{-1mm}
\begin{equation}
\langle {\cal E} | S^z_1\dots S^z_M|\F \rangle  =
\langle \E|\,
\prod_{k=1}^M 
( n_{j_k,\up}+  n_{j_k,\dwn}-1)/2 \,|\F \rangle
\label{n-correlators}
\end{equation}
\begin{equation}
\langle {\cal E} | S^-_i  S^+_j|\F \rangle =u_{ij}({\cal E},\F)=
\langle \E| 
c^{}_{i,\dwn} c^{}_{i,\up} 
c^\dagger_{j,\up} c^\dagger_{j,\dwn}
 \, |\F \rangle
\label{pairing-correlator}
\end{equation}
The vectors 
\nbeqa
\langle \E|&:=&\langle 0|\prod_{\alpha=1}^N S^-(e_\alpha)\, , \\|\F \rangle &:=& \prod_{\alpha=1}^N S^+(f_\alpha) |0 \rangle
\neeqa 
are exact $N$-pair eigenstates of (\ref{pairing})
(see Eqs.~(\ref{BA-state}), (\ref{gaudin-operators})).
Here,  we observe that  the evaluation of the CF's (like (\ref{n-correlators}),  (\ref{pairing-correlator}))
proceeds along  the action of local  operators, say $O_{loc}$,  onto the Bethe states,  the latter  involving  a collective reorganization of the vectors of the local 
Hilbert space ${\cal H}_{loc}$. Therefore, for any fixed Bethe root $u$, $O_{loc}$ has to be commuted with $B(u)$ to finally act on the vacuum $|0\rangle$.  
This gives rise to  a problem of combinatorial nature, whose solution 
is a non trivial task.  
In the next section we will look explicitely tame the combinatorics  for the 
special case of (\ref{n-correlators}),  (\ref{pairing-correlator}).

\section{Generating function for CF}
\label{ourwork}

In~\cite{SKLYANIN-CORRELATORS} Sklyanin suggested 
how the combinatoric complications involved in the calculation 
of the correlation functions  can be overcome resorting the Generating 
Function (GF) technique. 
He applied it~\cite{SKLYANIN-GAUDIN} to the $sl(2)$ Gaudin model~\cite{GAUDIN}.
The key role in his approach is played by a reordering 
making use of the Baker--Campbell--Hausdorff (BCH) formula 
for elements of the $SL(2)$ loop group associated with the 
Gaudin algebra ${\cal G}[sl(2)]$. 
In this section we exploit  the common algebraic root of the Gaudin 
and BCS models
to extend the Sklyanin theorem to the BCS model. 
The GF we will be looking at, is 
\begin{eqnarray}
C({\cal E}, {\cal H}, {\cal F}):=
\langle \E |\prod_{h \in {\cal H}}  S^z(h) |\F \rangle  
\label{general-correlators}
\end{eqnarray}
where the sets $\E, \F \subset \CC \setminus \E_0$ are 
(in general distinct) sets of solutions of the  Richardson equations~(\ref{re});
$\E_0:=\{2 \varepsilon_j, j=1\dots\Omega\}$; 
$\H \subset \CC \setminus (\E\cup\F\cup \E_0)$.  
The order of the correlation is the cardinality of $\H$: $|{\cal H}|$;
$|{\cal E}|$ and $|{\cal F}|$ are  fixed 
by the number of pairs $N$.
For instance, the one and two point CFs 
correspond to $|{\cal H}|=1$ and $|{\cal H}|=2$ respectively.    

Now we present the Sklyanin theorem for the GF of the $sl(2)$ 
Gaudin model and apply it to the BCS model.
Therefore we need the notation of 
the set of {\it coordinated partitions} 
${\cal P}= \{ P_l : l\in 1 \dots  |{\cal P}|\}$ 
of the sets ${\cal E}, \F, {\cal H}$ 
(see Ref.~\cite{SKLYANIN-CORRELATORS}):
the partition  $P\in \P$ is a set of triplets
$\{T_1\dots T_{|P|} \}$; the triplet $T\in P$ is  
$T=(\E_T,  \F_T, 
{\cal H}_T) $, where 
$\emptyset \neq {\cal E}_T \subset \E$, 
$\emptyset \neq {\cal F}_T \subset \F$ 
and ${\cal H}_T \subset {\cal H}$ 
such that $|\E_T|= |\F_T| >0, 
\quad |\H_T|\geq 0$. 
\\
The GF has  been evaluated  for the $sl(2)$ Gaudin 
model exploiting the BCH formula
for 
the  $SL(2)$ loop group generated by 
\nbeqa
S^-_{\phi(x)}&:=&\sum_{f\in \F}\phi_f S^-(f)\, ,\\
S^z_{\eta(x)}&:=&\sum_{h\in \H}
\eta_h S^z(h)\, ,\\
S^+_{\psi(x)}&:=&\sum_{e\in \E}\psi_e S^+(e)\, ,
\neeqa
where $\{S^z(u),S^\pm(u)\}\in \ga$ and  
$\phi(x),\, \eta(x),\, \psi(x)$ are meromorphic functions for  $x\in \CC$
with residues $\phi_f,\, \eta_h,\, \psi_e$ respectively~\cite{SKLYANIN-CORRELATORS}. 
This formula allows 
to rearrange the products between loop group elements 
in ~(\ref{general-correlators})\begin{multline*}
\langle\exp{S^-_{\phi(x)}}\exp{S^z_{\eta(x)}}\exp{S^+_{\psi(x)}} \rangle\\
=:\langle\exp{S^+_{\tilde{\psi}(x)}}\exp{S^z_{\tilde{\eta}(x)}}
\exp{S^-_{\tilde{\phi}(x)}} \rangle=
\langle\exp{S^z_{\tilde{\eta}(x)}}\rangle\; .
\end{multline*}  
Sklyanin proved the following theorem~\cite{SKLYANIN-NOTATION}.

{\bf Theorem}.
$C(\E,\H,\F)$
{\it is given by the formula
\begin{eqnarray}
\label{sklyanin-theorem}
\lefteqn{
C(\E,\H,\F)=(-1)^N } \\
&&\sum_{\P}\left(\prod_{T\in P}
n_{T}( |\E_T|)^{|\H_T|} 
S(\W_T\cup\H_T)\right) \prod_{h\in{\overline\H}_P} s(h) \nonumber 
\end{eqnarray}
where 
\begin{multline*}
S({\cal L})=1/2\pi i {\int}_\Gamma \; s(z)
\prod_{y\in {\cal L}} (z-y)^{-1}   dz\, ,\\
n_T:=-2 |\E_T|!\,(|\E_T|- 1)!\; ,\
\W_T:=\E_T\cup\F_T\; ,
\end{multline*}  
and ${\overline \H}_P:=
\H \setminus \bigcup_{T\in P} \H_T$. 
$C(\E,\H,\F)$ is a polynomial 
in $S$ with integer coefficients.} 
\\
Expression (\ref{sklyanin-theorem}) depends only on the sets 
$\W:=\E\cup\F$ and $\H$~\cite{SKLYANIN-CORRELATORS,KOREPIN-GAUDIN}; 
for the Gaudin model $\W$ is a set of 
solutions of ~(\ref{re}) for $g \to \infty$;
for the BCS model $\W$ is a set of solutions of the  
Richardson equations~(\ref{re}) for generic $g$.  
The scalar products of Bethe states (and their norms)
are a corollary of the Sklyanin theorem~(\ref{sklyanin-theorem}) 
for $\H=\emptyset$:  
$\langle \E | \F \rangle=C(\E,\emptyset,\F)$.
Its consent with the determinant 
formulas\cite{GAUDIN,RICHARDSON-CORRELATION} has been elucidated in 
Refs.\cite{SKLYANIN-CORRELATORS,KOREPIN-GAUDIN}.
\\
We point out that the GF~(\ref{general-correlators}) 
has simple poles in the set $\E_0$. This will play 
a key role in the following.


{\it Correlation functions}.
The charge and pairing CFs
are matrix elements of the $su(2)$ Lie algebra (instead of elements of $\ga$) 
using vector states of $\ga$. 
The projection from the $sl(2)$ loop algebra on its Lie algebra 
is performed by taking the residue of $C(\E,\H,\F)$ in the poles 
$h_l=2\varepsilon_{j_l}$ 
for $h_l\in \H$, $l \in \{ 1 \dots M\}$.
The charge CFs~(\ref{n-correlators}) are 
\begin{equation}
\langle {\cal E} | S^z_1\dots S^z_M|\F \rangle= 
\lim_{\H \to \E_0 } (\H-\E_0)C(\E,\H,\F) 
\label{charge-residue}
\end{equation}
where $\H \to \E_0$ involve $ h_l \to 2 \varepsilon_{j_l}$ 
$\forall l$ and $\H-\E_0$ means 
$\prod_l (h_l -2 \varepsilon_{j_l})$. 
Using (\ref{sklyanin-theorem}) yields 
\begin{eqnarray}
\label{n-charge-correlators}
&&\langle {\cal E} | S^z_1\dots S^z_M|\F \rangle =(-1)^N \prod_l^{M}s_{j_l}   \\
&&\sum_{P \in \P_1}  \left ( \prod_{T\in T_0} n_{T} 
S(\W_T)  \right )
\left (\prod_{T\in T_1 \atop \H_T = h_T}  
  \frac{ n_{T}|\E_T| }
{\prod\limits_{y\in \W_T} (h_T -y)}\right) 
\nonumber
\end{eqnarray}
where  $\P_k:= \{P\in \P : \max\limits_{T\in P} |H_T| = k \}$;
$T_k:= \{T\in P : |H_T| = k \}$. The quantity $S(\W_T)$ is
\begin{eqnarray}
&&S(\W_T)= \sum_{e \in \W_T} 
\frac{s(e)}{\prod\limits_{x\in \W_T \atop x \neq e} (e -x)} - \\ 
&&\sum_{d \in \W_T}\left (\frac{s(d)}{\prod\limits_{x\in \W_T \atop x \neq d} (d -x)}
\sum_{y \in \W_T \atop y \neq d }\frac{1}{(d -y)}+ \frac{s^{[2]}(d)}
{\prod\limits_{x\in \W_T \atop x \neq d} (d -x)}\right ) \nonumber    
\end{eqnarray}
where $e$ and $d$ are elements 
appearing singly and doubly in $\W_T$ respectively. 
The pairing CF~(\ref{pairing-correlator}) can be  
extracted  from  $C(\tilde{\E},\emptyset,\tilde{\F})$ where   
the vectors in ~(\ref{general-correlators}) are 
$\langle \tilde{\E}|:= \langle \E| S^-(z_1)$ and 
$|\tilde{\F}\rangle :=S^+(z_2) |\F\rangle $.
Then $u_{lm}(\E,\F)$ is   
\begin{eqnarray}
u_{lm}(\E,\F) =\!\!\!\lim_{\scriptstyle z_1\to 2 \varepsilon_l 
\atop \scriptstyle z_2\to 2 \varepsilon_m}\!\!\!( z_1- 2 \varepsilon_l) 
( z_2- 2 \varepsilon_m)  C(\tilde{\E},\emptyset ,\tilde{\F})
\label{limit-pairing}
\end{eqnarray}
$C(\tilde{\E},\emptyset ,\tilde{\F})$ is then calculated using 
the Sklyanin theorem.
For $l\neq m$ formula (\ref{limit-pairing})  gives
\begin{eqnarray}
\label{problem}
&&u_{lm}({\cal E},\F)  =(-1)^{N+1} \\ 
&&\sum_{P \in \tilde{\P}_1}  \left (\prod_{T\in \tilde{T}_0} n_{T} 
S(\W_T) \right ) 
\left (\prod_{T\in \tilde{T}_1} 
  \frac{n_{T}  s_{l_T}}
{\prod\limits_{y\in \W_T} 
(2 \varepsilon_{l_T} -y) }\right) 
\nonumber
\end{eqnarray}
where $Z:=\{z_1,z_2\}$ and $l_T$ is one of $l$ and $m$;  
\nbeqa
\tilde{\P}_k&:= &\{P\in \P : \max\limits_{T\in P} |Z_T| = k \}\; ,
\\{T}_k&:=& \{T\in P : |Z_T| = k \}\; .
\neeqa
The pairing CF for 
$l=m$ can be obtained by 
a variation  of the  procedure 
depicted above:
\nbeq
\lim_{\scriptstyle z_1\to 2 \varepsilon_l 
\atop \scriptstyle z_2\to 2 \varepsilon_l}( z_1- 2 \varepsilon_l) 
( z_2- 2 \varepsilon_l)  C(\tilde{\E},\emptyset ,\tilde{\F})\;.
\neeq
But in the present case ($s_j=-1/2\, \forall j$) 
it is more convenient employing the formula~(\ref{n-charge-correlators}) 
because
$S^\pm_j S^\mp_j=1/2 \pm S^z_j$. 

We comment that practical use of the formulas 
is limited by the vastly increasing number of partitions, which
depends on the number of pairs $|\E|=N$ and the order of the 
CF $|\H|$. We want to emphasize that no complete knowledge
of all the eigenstates is required. It doesn't show 
any dependence of the Hilbert space dimension either.
We finally point out that the results apply to arbitrary $s_j$ 
(i.e. any degeneracy of the single particle levels).

\section{Determinant representation of the Correlation Functions}
\label{ZhouetalWork}
In this section we will sketch how the formula  obtained 
above can be simplified by recasting the CF's into sums of  determinants of certain $N\times N$ matrices. 
As it was remarked above the  scalar products between  the BCS Bethe states 
$
 \langle \E | \F \rangle=C(\E,\emptyset,\F)
$ 
can be  expressed as determinants. Within the formalism we exploited in the previous section (see Eq.(\ref{sklyanin-theorem})) in fact

 \begin{multline}
 C(\E,\emptyset,\F)=(-1)^N  \\
 \sum_{\P}\left(\prod_{T\in P}
n_{T}( |\E_T|)^{|\H_T|} 
S(\E_T\cup \F_T)\right)
\; ,
 \end{multline}
 where $S(\E\cup \F)$ can be written as
 \begin{equation}
 S(\E\cup \F)=\sum_{e\in \E} \sum_{f\in\F} \frac{S(e,f)}{\prod_{e'\neq e}\prod_{f'\neq f}(e-e')(f-f')}\;. 
 \end{equation}
 Sklyanin proved\cite{SKLYANIN-GAUDIN} that $C(\E,\emptyset,\F)$
 can be indeed written as a polynomial that is linear in each of 
$\displaystyle{S (e, f)=
\frac{s(e)-s(f)}{e-f}
}$,   $e\in \E,f\in \F$  (see Eq. \ref{re}). Consistently with Richardson's old result~\cite{RICHARDSON-CORRELATION}, it can be expressed as a sum of $N!$ determinants
\beq
\label{scalar-sklyanin}
C(\E,\emptyset,\F)=\sum_{\pi \in S_N } {\cal M}_\pi \; ,
\eeq 
where $\pi$ is an element of the symmetric group $S_N$ and $M_\pi$ is defined as
\beqa
&&(M_\pi)_{\alpha \alpha}=S(e_\alpha, f_{\pi(\beta)})+ 2\sum_{\alpha,\alpha'} \frac{1}{(e_\alpha-e_\alpha')(f_{\pi(\alpha)}-f_{\pi(\alpha')})} \;, \nonumber \\
&&(M_\pi)_{\alpha \beta}=- \frac{2}{(e_\alpha-e_\beta)(f_{\pi(\alpha)}-f_{\pi(\beta)})} \;. \nonumber
\eeqa
A major simplification was achieved by Slavnov who was able to express the scalar product as a single determinant \cite{slavnov}. 
Therefore the  Eq.(\ref{scalar-sklyanin}) can be recast into
\beqa
\langle{\cal E}| {\cal F}\rangle&=&
\frac {\prod\limits^N_{\b=1} \prod\limits^N_{\a \neq \b}
(f_\b-e_\a)}
{\prod\limits_{\b <\a} (e_\b -e_\a) \prod\limits_{\a <\b} (f_\b -f_\a)}  \no\\
&& \quad {\rm det}_N H(\{ f_\a \}, \{ e_\b \}),\label{slavnov_BCS}
\eeqa
As discussed in Sect.\ref{model}  the CF's  of 
the BCS model are identical to those of the Gaudin model,
with the para\-meters $e \;,\; f $ satisfying Richardson's equations
(\ref{re}) instead of the Gaudin-equations. 
The entries of the $N \times N$ matrix $H(\{ f_\a \}, \{ e_\b \})$
are
\beqa
H_{ab} &=& \frac {f_b -e_b}{f_a -e_b} 
\left ( \sum ^\Omega _{j=1} \frac {1}
{ (f_a -\e _j)(e_b -\e _j)} \right. \no\\
&& \left. -2\sum _{\a \neq a} \frac {1}{(f_a
-f_\a)(e_b -f_\a)} \right ).
\label{slavnov-mat}
\eeqa
The norms of the states are obtained for  $e_\a \rightarrow f_\a$ in (\ref{slavnov_BCS}) and give
$|\psi(e_\alpha)|^2=det_N G_N$ where $G$ is the Gaudin matrix given by
\beq\label{GaudinMat}
G_{ab}=\left\{\displaystyle \begin{array}{lcc}
  \displaystyle\frac{2}{(e_a-e_b)^2}& ; & a\neq b\\
  \displaystyle \sum\limits_i^{}  \frac{2}{(e_a-\eps_i)^2}-\sum_{c\neq a} \frac{2}{(e_a-e_c)^2}
  & ; & b=a  \end{array} \right .
\eeq
The various stages of the calculation of CF's proceed through certain 
recurrence formulas involving (\ref{slavnov_BCS}) as a basic ingredient 
(see \cite{KOREPIN-BOOK} for the details). Therefore  the CF's result to be  determinants as well.
We comment that such a simplification was first achieved   after a {\it tour de force} on integrable spin $1/2$ theories (beyond the quasi-classical 
expansion) leading to the so called solution of the inverse problem~\cite{kmt98,mtgk}.
The main accomplishment  is that the  lattice spin variables are  expressed in terms of the entries of the monodromy matrix $A(u|\epsilon)$,
$B(u|\epsilon)$,  $C(u|\epsilon)$,  $D(u|\epsilon)$ in a closed (and simple) form. This allows to evaluate the CF's in the non-local Hilbert space spanned  by the Bethe 
vectors (instead of expressing the Bethe states in ${\cal H}_{loc} $).  
Zhou {\rm et al.}~\cite{LINKS}  calculated the relevant quantities for the BCS model  through the 
quasi-classical limit~\eqref{transf}, generalizing the solution of the inverse problem to non-fundamental integrable spin theories 
(where the auxiliary and the quantum spaces have different dimensions).  
The formulae read~\cite{LINKS}\bea
S^-_i &=& \prod^{i-1}_{\a=1} t(\e_\a) K^{-i+1}B(\e_i) K^{i-1}
\prod ^i_{\a=1} t^{-1}(\e_\a),\no\\ 
S^+_i &=& \prod^{i-1}_{\a=1} t(\e_\a) K^{-i+1}C(\e_i) K^{i-1}
\prod ^i_{\a=1} t^{-1}(\e_\a),\no\\
S^z_i &=& \prod^{i-1}_{\a=1} t(\e_\a) K^{-i+1} 
\frac {(A(\e_i)-D(\e_i))}{2} K^{i-1}
\prod ^i_{\a=1} t^{-1}(\e_\a), \no 
\eea
with $ K := 
\exp (-2\eta \sum^\Omega_{j=1} S^z_j/g\Omega)$ 
being essentially the total $S^z$, and $t(u)$ being the transfer matrix. 
The form factors are obtained as~\cite{kmt98}
\bea
&& _{N+1}\langle{\cal E}|S^-_m| {\cal F}\rangle _{N} = \no\\
&& \frac {\prod ^{N+1}_{\b=1} (e_\b - \e_m)} 
{\prod ^N_{\a=1} (f_\a - \e_m)} 
\frac { {\rm det}_{N+1} \T (m, \{ e_\b \}, \{ f_\a \})}
{\prod _{\b > \a} (e_\b -e_\a) \prod _{\b <\a} (f_\b -f_\a)},\label{S-ff}\\
&& \langle e_1,\cdots,e_N|S^z_m|f_1,\cdots,f_N\rangle = 
\prod ^N_{\a=1} \frac {(e_\a - \e_m)} 
{(f_\a - \e_m)}\no\\ 
&&~\frac { {\rm det}_N \left (\frac {1}{2} \TT(\{ e_\b \}, \{ f_\a \}) 
- Q (m, \{ e_\b \}, \{ f_\a \}) \right )} 
{\prod _{\b > \a} (e_\b -e_\a) \prod _{\b <\a} (f_\b -f_\a)},
\label{Szff}\eea
where $ _{N+1}\langle{\cal E}| $ and $| {\cal F}\rangle _{N}  $ indicate   Bethe states with $N+1$ and $N$ rapidities respectively. The matrix elements of $\T$ and $Q$ given by
\bea
\T_{ab}(m) =&& 
\prod ^{N+1}_{\stackrel {\a=1}{\a \neq a}} (e_\a - f_b)
\left ( \sum ^\Omega _{j=1} \frac {1}
{ (f_b -\e _j)(e_a -\e _j)} \right. \no\\ 
&& \left. -2\sum _{\a \neq a} \frac {1}{(f_b-
e_\a)(e_a -e_\a)} \right ),
~~~b < N+1, \no\\
\T_{aN+1}(m)  && =  \frac {1}{(e_a -\e_m)^2}, \ \
Q_{ab}(m) = \frac {\prod _{\a \neq b} (f_\a-f_b)} {(e_a-\e_m)^2}.
\no \eea
$\TT$ is the $N\times N$ matrix 
obtained from $\T$ by deleting the last row and column and replacing $N+1$ by
$N$ in the matrix elements. Here, we assume that both $\{ f_\a \}$ and
$\{ e_b \}$ are solutions to Richardson's Bethe equations
(\ref{re}). 

The two-point correlation functions are
\bea
&&\langle{\cal E}|S^-_mS^+_n|{\cal F}\rangle = \no\\
&& \sum _{\a=1}^N \frac {1}{f_\a -\e_n}
\langle e_1,\cdots,e_N|S^-_m|f_1,\cdots,\hat  f_\a, \cdots, f_N\rangle -\no\\
&& \sum _{\a \neq \b} \frac {1}{(f_\a -\e_n)(f_\b -\e_n)} \no\\
&&\langle e_1,\cdots,e_N|S^-_mS^-_n|f_1,\cdots,\hat f _\a,
\cdots, \hat f _\b, \cdots, f_N\rangle. \label{S+S-ff}
\eea
Here, the hat denotes that the corresponding parameter is not present in
the set. Since $\{ e_a \}$ is a solution of the Bethe equations, 
$\langle{\cal E}|S^-_m|f_1,\cdots,\hat f _\a, \cdots, f_N\rangle $
is the form factor given before, while
\begin{multline} \label{2ff}
{}_{N}\langle{\cal E} |S^-_mS^-_n|{\cal F}\rangle_{N-2}=\\
\frac {\prod\limits ^N_{\b=1} (e_\b - \e_m)(e_\b-\e_n)} 
{\prod\limits ^{N-2}_{\a=1} (f_\a - \e_m)(f_\a -\e_n)}
 \frac { {\rm det}_N \bar{\T} (m,n, \{ e_\b \}, \{ f_\a \})}
{\prod\limits_{\b > \a \atop \gamma < \delta} (e_\b -e_\a)(f_\gamma -f_\delta) }
\end{multline} 
with
\nbeqa
\bar{\T}_{ab}(m,n) =&& 
\prod ^N_{\stackrel {\a=1}{\a \neq a}} (e_\a - f_b)
\left ( \sum ^\Omega _{j=1} \frac {1}
{ (f_b -\e _j)(e_a -\e _j)} \right. \\ 
&& \left. -2\sum _{\a \neq a} \frac {1}{(f_b-
e_\a)(e_a -e_\a)} \right ),
~~~b < N-1, 
\neeqa
\nbeqa
\bar{\T}_{aN-1}(m,n)  && =  \frac {2e_a-\e_m-\e_n}{[(e_a -\e_m)(e_a-\e_n)]^2},\\ 
\bar{\T}_{aN}(m,n)  && =  \frac {1}{(e_a -\e_m)^2}, 
\neeqa
In (\ref{2ff}) $m \neq n$ is assumed, and it is zero if $m=n$. 
	
The expression for the charge CF can be obtained from the form factors and CF above\cite{calabrese_static}:
\begin{eqnarray}\label{szszff}
&&  \langle{\cal E} |S^z_mS^z_n|{\cal F}\rangle
= {{{\langle{\cal E} |{\cal F}\rangle} }\over{4}}  +  \\
&&\sum_{\alpha=1}^{N}  \langle{\cal E}|  \frac{ S^-_m}{2(\e_m f_\alpha)} + 
\frac{ S^-_n}{2(\e_n f_\alpha)}   |f_1,\cdots,\hat f _\a, \cdots, f_N\rangle +
\nonumber \\ && 
\sum_{\beta \ne \alpha}^{N}  { \langle{\cal E}} |  \frac{ S^-_m S^-_n}
{(\e_m-f_{\alpha})(\e_n-f_{\beta})}|f_1,\cdots,\hat f _\a,
\cdots, \hat f _\b, \cdots, f_N\rangle \nonumber
\,.
\end{eqnarray}
The static CF of interest here are obtained with $ \langle{\cal E} |=
\langle{\cal F} |$.

\subsection{Reduction formulas}
The last significant progress for the evaluation of the CF for  $ \langle{\cal E} |= \langle{\cal F} |$  was pursued by Faribault, Calabrese and Caux\cite{calabrese_static}.
They managed to reduce the complexity of the above expressions to sums over only $N$ determinants.

Both  $\langle S_m^- S_n^+\rangle$ and $\langle S_m^z S_n^z\rangle$ involve the evaluation of the form factors 
(\ref{S-ff})
and (\ref{2ff}) for  $e\to f$. In such a limit the CF are still a sum of two terms, each one involving sums of $N$ determinants of modified Gaudin matrices. In Ref. \cite{calabrese_static} the specific symmetry of the Richardson equations was exploited to reduce the CF to  a {\it single}  term expressed as a sum of $N$ determinants
(see \cite{calabrese_static} for the detailed calculations).
In this way, the final result for the correlation function is
\be
\langle {\cal E} |S_m^- S_n^+ | {\cal E}  \rangle= \sum_{q=1}^{N}
\frac{e_q-\e_m}{e_q-\e_n} D_q^{(m,n)} \,.
\label{S+S-red}
\ee
where
 \be
D_{q,i}^{(m,n)} =
\begin{cases}\displaystyle
\vec{G}_i- \frac{K_{iq}}{K_{i+1q}} \vec{G}_{i+1} \ &
i<q-1,\\ \displaystyle
\vec{G}_{i}+2\frac{ (e_q-\e_n) (e_{q-1}-\e_m)}{e_{q-1}-e_q} \vec{B} \  
&
i=q-1,\\
\vec{C} \  &
i=q,
\\ \vec{G}_i& i> q.
\end{cases}
\ee
Here, 
\bea
C_a &=& \frac1{(w_a-\e_m)^2}\,, \label{vecC}\\
B_a &=& \frac{2w_a-\e_m-\e_n}{(w_a-\e_m)^2(w_a-\e_n)^2}\,.
\label{vecB}
\eea
Similarly 
$\langle S_m^z S_n^z\rangle$ is given by~\cite{calabrese_static}
\begin{multline}
\langle {\cal E} |S_m^z S_n^z | {\cal E}  \rangle =  \\
  \frac{\langle {\cal E}|{\cal E} \rangle}{4} -
\frac12 \sum_{q=1}^{N} (\det D_q^{(m,n)}+\det D_q^{(n,m)})\,.
\end{multline}
The formulas obtained above for the correlation functions are completely  
general
and are valid for any choice of the Hamiltonian parameters $\e_\a$ and  
$g$
(with some caveat in the limit of coinciding energies). 
The low level of complexity of this representation as sum of $N$  
determinants
of $N$ by $N$ matrices allows one to have access to the static  
correlation functions for systems with a reasonable number of pairs.

\section{Canonical pairing fluctuations}
\label{seccorr}
\begin{figure}[t]
\vspace{5mm}
\includegraphics[width=0.9\columnwidth]{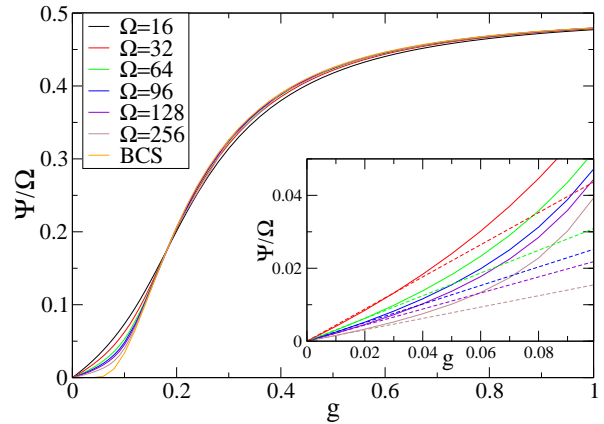}
\caption{(Color online) Canonical order parameter $\Psi$ as a function  
of
$g$ for different numbers of pairs $\Omega=N/2$. The  BCS result $\Psi=\Omega/(4 g \sinh (1/2g))$
is plotted for comparison. Inset: Small g behavior of $\Psi$ compared with large N expansion. With kind permission by authors of \cite{calabrese_static}.}
\label{figPsi}
\end{figure}

\begin{figure}[t]
\includegraphics[width=\columnwidth]{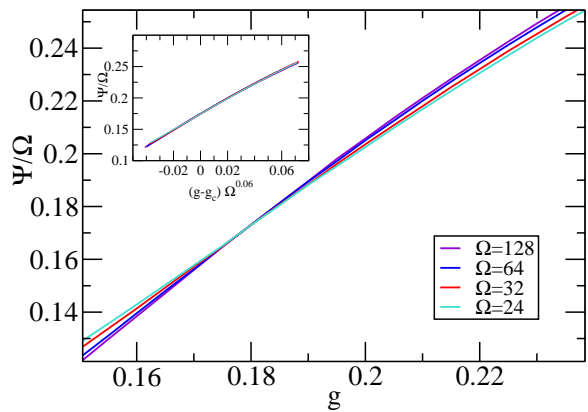} \vspace{2mm}
\caption{(Color online) The data collapse for $\eta=1.00\pm 0.001$ and $\nu=16.\bar{6}$. The data points have been read out of the figure 8 taken from Ref.~\cite{calabrese_static}.}
\label{Collasso}
\end{figure}

In the canonical ensemble the  conventional BCS  order parameter  $\Delta={}_N\langle {\cal U} |c^\dagger_\downarrow c^\dagger_\downarrow| {\cal U}\rangle_N $
is vanishing exactly.
Nevertheless the pairing instability can be characterized by studying the correlation function
\begin{multline}
u_i v_i :=  \langle S^-_i S^+_i \rangle \langle S^+_i 
S^-_i\rangle \\
=\langle c_{i\up }^\dagger c_{i \dwn}^\dagger 
c^{}_{i \dwn} c^{}_{i\up}  \rangle - \langle c_{i\up }^\dagger c^{}_{i\up}  \rangle 
\langle c_{i \dwn}^\dagger  c^{}_{i \dwn} \rangle
\end{multline} 
indicating the tendency that electrons form Cooper pairs  
instead of uncorrelated electrons. The  canonical BCS 
order parameter is\cite{vondelft,MASTELLONE}
\be
\Psi=\sum_{i=1}^\Omega u_i v_j\, ,
\ee
 which in the limit of large volume $\Omega$ and large $N$ reduces  
to the BCS value (see Sect.\ref{thermo}).
 We observe that, in contrast with a normal Fermi gas, a system 
with pairing instability will take an energetic advantage  
by increasing $\Omega$ for fixed $\Omega/N$  
(because the phase space available for coherence is enlarged).
Therefore, energy correlations are short ranged in a normal Fermi gas 
and long ranged in the presence of a 
pairing coherence. 
Accordingly, the footprint for an ongoing pairing instability is 
a finite size scaling ansatz
\begin{equation}
\Psi=\Omega^\eta F\left [ (g-g_c) \Omega^{1/\nu}\right ]
\; .
\end{equation}
\begin{figure}[t]
\includegraphics[width=\columnwidth]{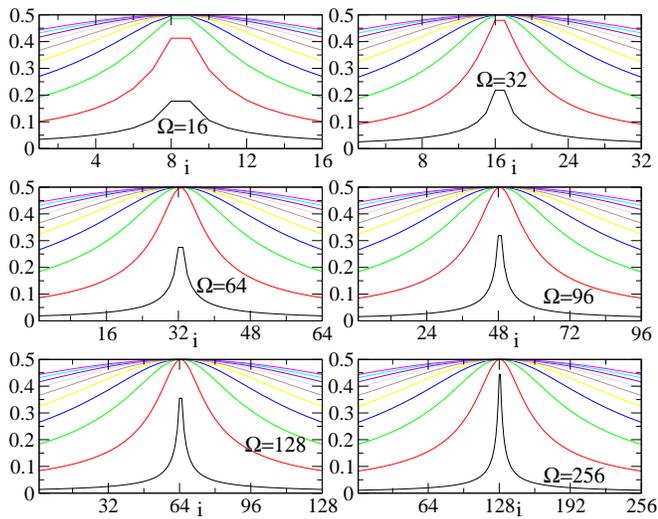} \vspace{2mm}
\caption{(Color online) Pair distribution $u_i v_i$ as function of $i$. 
Each plot is for fixed $\Omega$ but for several different $g$ going from $0.1$ to
$1$ in steps of $0.1$. With kind permission by the authors of \cite{calabrese_static}.}
\label{pair-N}
\end{figure}
\begin{figure}[t]
\includegraphics[ width=\columnwidth]{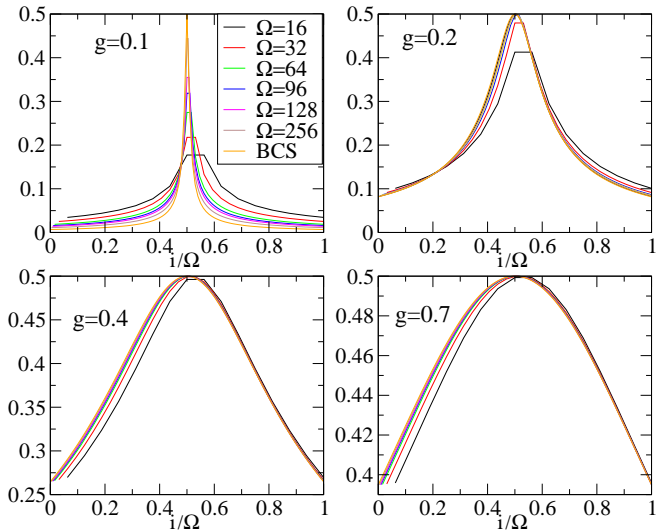} \vspace{2mm}
\caption{(Color online) Pair distribution $u_i v_i$ as function of $i/\Omega$ for different values of $\Omega$. The four plots are taken at variable
$g$. With kind permission by the authors of \cite{calabrese_static}.}\label{pair-g}
\end{figure}
The above quantity was evaluated exploiting the exact formulas 
(\ref{n-charge-correlators}) and  (\ref{Szff}). 
The rapidities involved in the equations at finite size are obtained by 
solving the Richardson equations numerically for the model parameters, 
which here are equally spaced single particle energy levels $\epsilon_i= i$ 
and half filling $\Omega=2 N$ 
(see Ref.\cite{numericsRichardson} for the details).
The results for $u_i v_i$ are shown in Figs. \ref{pair-N}
and \ref{pair-g}. Whereas the former shows the $g$ dependence at fixed $N$,
in the latter each plot consists of the various curves
at fixed $g$ (=0.1, 0.2, 0.4, 0.7) for varying $N$.
It is clear that the results tend to the BCS result. 
In Ref.\cite{calabrese_static}   was noticed that this convergence is the slower, the smaller $g$ is:
for $g=0.1$ the maximum at $N=256$ is only 90\% close to the 
asymptotic result wheras at $g=0.7$ the $N=16$ result is already at 99.8\%.
In Fig. \ref{figPsi} the order parameter $\Psi/\Omega$ as a  
function of
$g$ for several values of $N$ is shown and compared with the BCS result:
for $N=128$, $\Psi$ is almost
indistinguishable from its limiting value for large enough $g$.
\begin{figure}
\includegraphics[width=\columnwidth]{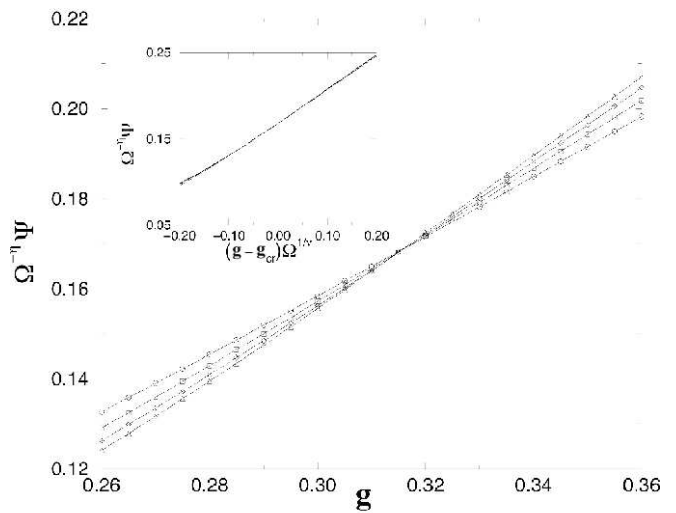}
\caption{Scaling of $\Psi$.  
Data for $\Psi$, as a function of $\alpha  = g/\delta$ and 
 $\Omega= 2 N = even$, which show clearly a phase transition ($N = 8 (circles), N = 12 (squares), N = 16 (diamonds), N = 20 (triangles)$ ). 
 In the inset it is shown the data collapse. The parameters are
$g^*_1=0.315$, 
$\eta(g^*_1)=0.940$, and $1/\nu=0.26$. Taken from~\cite{MASTELLONE}.}
\label{scaling_andrea}
\end{figure}
\begin{figure}
\includegraphics[angle=0,width=\columnwidth]{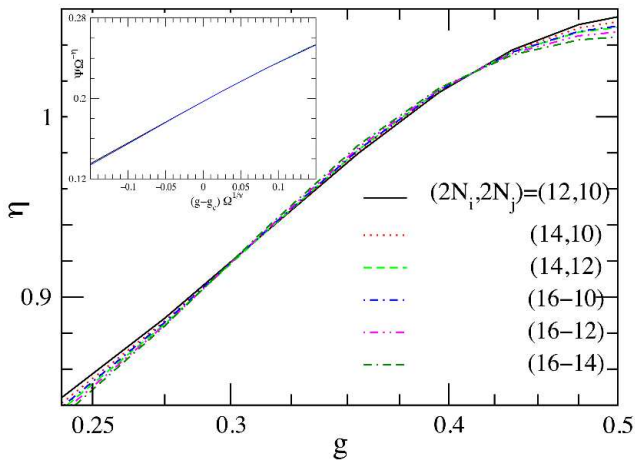}
\caption{The figure shows two crossing points: the first
is in agreement with Ref.~\cite{MASTELLONE}; the second is at 
$g^*=0.417$, $\eta=1.028$. The data collapse is seen in the inset
for $1/\nu=0.15$. Taken from~\cite{amico-osterloh}.}
\label{scaling}
\end{figure}
The scaling of $\Psi$ was originally obtained  in \cite{MASTELLONE} for small size at a given value of the pairing coupling  $g_1$ (see the caption of Fig.\ref{scaling_andrea}). 
In Ref.\cite{amico-osterloh}  a further scaling point $g_2$ was evidenced (see the caption of Fig.\ref{scaling}).
In order to extract the finite-size scaling, 
$\log [ \Psi_\Omega(g)/\Psi_{\Omega'}(g)]/\log[\Omega/\Omega']$ 
was taken in consideration for different valus for $\Omega$ and $\Omega'$.
At a scaling point all these curves cross 
($\eta(\Omega , \Omega',g*)\equiv \eta(g*)$) as shown in Fig.~\ref{scaling}.
The physical meaning of two apparent ``scaling points'' was unclear and
deserved further analysis. This analyis was significantly extended in \cite{calabrese_static}, but without any scaling analysis.
However, there is no second scaling point visible in their analysis 
for larger pair number ($N\geq 32$).
In figure \ref{Collasso} we present the data collapse of the data of 
Ref.~\cite{calabrese_static}. The value for $\eta$ is in accordance with $\eta=1$ with an error about $0.01$. The second coefficient $\nu=16.\bar{6}$ leads to the searched-for data collapse.

It is instructive that the computed leading term of $\Psi$
for large $\Omega$ and $g$
\beq
\frac{\Psi}{N}=\frac{1}{2}-\frac{1}{48 g^2}+O(g^{-3},N^{-1})
\eeq
coincides with that of Ref.~\onlinecite{sild-01}. 
Instead, for small $g$ and large $N$ \be
\frac\Psi{N}=
g\frac{\ln (3+\sqrt8))}{\sqrt{N}}+O(1/\ln N)\,,\quad {\rm for}\;  
g\ll1\,.
\ee
The $N$ independent result for large $g$ and the 
$N^{-1/2}$ dependence at small $g$ gives a hint towards the non-perturbative 
nature of superconductivity. Further studies on the finite size corrections of the BCS pairing amplitude 
were performed in\cite{finitesize_altshuler}.

We close the discussion on the canonical pairing fluctuations 
mentioning the relation between $\Psi$ and the 
Onsager-Penrose-Yang parameter~\cite{ODLRO} for the long-range 
off-diagonal order$\Psi_{OD}$,
taking into account the effect of non-diagonal correlations.\\
{}
Although 
$\Psi_{OD}$ can be obtained with the formulas for $\expect{S^+_iS^-_j}$, 
a much easier route is to apply the Hellmann-Feynman theorem:\be
\Psi_{OD}=
\frac{1}{N} \sum^N_{i,j=1} \left< S^+_i S^-_j\right>=
- \frac{1}{N} {\partial E_0(g)}/{\partial g}\; ,
\label{FVth}
\ee
where $E_0$ is the ground state energy of the BCS model.
In the thermodynamic limit
$\Psi_{OD}^{N=\infty}=\frac{\Psi^2_{N=\infty}}{N}\,$. 
However, the two quantities are independent for finite sizes.
Tian \rm{et al.} ~\cite{ineq} proved that $\Psi$ and $\Psi_{OD}$
satisfy the following relations for any value of $g$ and $N$
\be
\frac{1}{N} \Psi(\Psi-1)\leq \Psi_{OD}\leq1+\frac{\Omega}{N} \Psi\,.
\ee
For $N\to\infty$ these are trivial bounds, but not so for finite $N$.

\section{Thermodynamic limit}\label{thermo}
The Richardson equations ~\eqref{re} admit an electrostatic analogy~\cite{RICHARDSON,GAUDIN,roman02,admor02}, where the eigenenergies $\eps_i$ 
and solutions $e_\alpha$
both are interpreted as point charges
of the strengths $-1/2$ and $1$ respectively. The thermodynamic limit is 
performed making use of this analogy.\\ 
Define
\beqa 
\rho(x_j)&=&\frac{1/2}{\Omega(x_{j+1}-x_j)} \\
\sigma(z_\alpha)&=& \frac{1}{\Omega |e_{\alpha+1} - e_\alpha|}\\
g&\to& \frac{g}{\Omega}\; .
\eeqa
This choice leaves the Debeye-shell invariant.
Inserting this into the Richardson equation~\eqref{re}, $\Omega$
cancels and we obtain
\beq
\frac{1}{g} + \int {\rm d}\eps \frac{\rho(\eps)}{z-\eps}
- \int |{\rm d}z'| \frac{\sigma(z')}{z-z'}=0\; .
\eeq
\begin{figure}[t]
\includegraphics[width=\columnwidth]{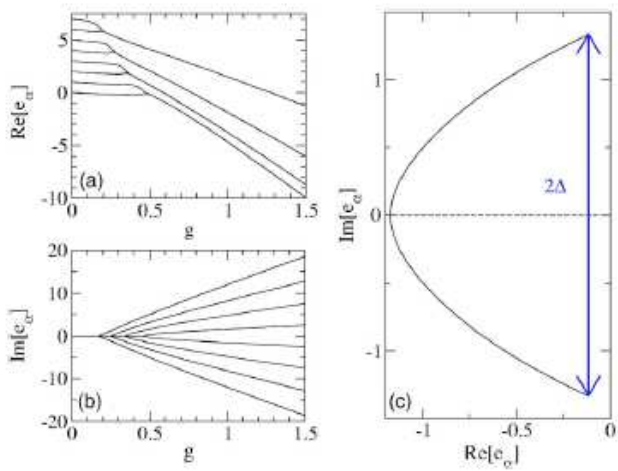} \vspace{2mm}
\caption{(Color online) {\it Left:} Solutions to the Richardson eqation~\ref{re}. {\it Right:} A picture on how the BCS-gap is determined.}
\label{Arc}
\end{figure}
Following the works ~\cite{RICHARDSON,GAUDIN} 
the BCS-gap $\Delta_{BCS}=2\Delta$ is the imaginary opening of the arc 
solving the Richardson equations, $a=\lambda\pm i \Delta$ (see Fig. \ref{Arc}), 
and the gap equation becomes
\beqa
\frac{1}{g}&=&\sum_i \frac{d_i}{\sqrt{(\eps_i-\lambda)^2+\Delta^2}}\\
&\to& \int {\rm d}\eps\ \frac{d(\eps)\rho(\eps)}{\sqrt{(\eps-\lambda)^2+\Delta^2}}\; ,
\eeqa
where $d_i$ ($d(\eps)$) is the multiplicity of the level $\eps_i$.
This resuls in the known expression $\Delta=\frac{\omega_D}{\sinh 1/2g}$ for
the equally spaced model.
A good summary of Gaudin's article together with a modern numerical analysis of
the solutions for the BCS-model is found in~\cite{roman02}.
The generalization of the electrostatic analogy to more general settings 
(the trigonometric and hyperbolic BCS-model) can be seen in~\cite{admor02}. 

\section{Further directions}
In this article we have reviewed the current understandings of the mesoscopic fluctuations of the pairing instability based on Bethe ansatz techniques. 
The relevant quantity is a correlation function (CF), where the physical observables are evaluated as a {\it static} expectation value in the eigenstates of the Hamiltonian.    
The progress in the field of exact solutions are mature enough to allow an exhaustive analyis of superconductivity from mesoscopic regimes to thermodynamic limit at 
equilibrium.
An important piece of information, however, comes from the study of the 
system out of equilibrium. The typical picture is provided by 
transport experiments were the 
{\it dynamical} CF, $G(t)= \langle O(0) O (t) \rangle$,  
are the interesting quantities to be calculated. The formula 
for  $G(t)$ involve an additional level of complexity. 
The basic ingredients are off diagonal correlations, namely static CF 
between different eigenstates.
The first exact off-diagonal CF  for the BCS-model obtained in 
~\cite{amico-osterloh} could only be calculated for very small sizes.
The better performance of the determinant expression in ~\cite{LINKS} 
could be even further improved in 
~\cite{calabrese_static} to make reasonably higher pair numbers accessible.
Nevertheless,  this is not the end of the story because in principle 
{\it all} the off-diagonal CFs are involved in $G(t)$.
Fortunately, the problem can be simplified for the BCS model because the eigenstates do not contain the coupling constant {\it explicitly}.
In a very relevant paper, Faribault, Calabrese and Caux combined 
numerical and analytical analysis to realize that indeed only a 
relatively small amount of excitations contribute 
significantly to the dynamical CF~\cite{calabrese_dynamics}. 
As a consistency check they used exact sum rules relating the dynamical 
to static CF (the latter can be accessed easily). 
They discovered that the weight of the multi-particle excitations 
is suppressed increasing $N$: in the thermodynamic limit the two-particle 
excitations are hence dominant in the calculation of $G(t)$~\cite{abacus}. 
This is the ultimate reason why the Bogoliubov mean-field results (just neglecting the higher order correlations) coincide 
with the exact ones in the thermodynamic limit.
 
An important problem that has been intensely studied in the recent literature 
is the response of a given system when pushed out of equilibrium 
by a sudden change in some control parameter: the quantum quench
(see \cite{quench} for a review). 
The richness and complexity of this problem is very much related to 
the developing nonlocal character of the correlations in the system 
by time evolution~\cite{calabrese-cardy}. 
Remarkably, this kind of issues can be explored experimentally at
the quantum level by realizing highly controllable 
quantum many-particle systems with cold atoms~\cite{cold-atoms_experimental}. 
We  observe, however, that pairing fluctuations in cold atoms are expected 
to be more evident in transport experiments rather than in the 
popular expansion protocols (where the increase of single 
particle kinetic energy might mask the crossover). 
Arrays of coupled microcavities are potentially interesting alternative 
experimental platforms~\cite{photonics,coupled-cavity}.
Those problems are studied through the dynamical CF as well. 
The time evolution starts, because the eigen-basis where the 
wave function of the system lives, changes after the quantum quench. 
The computational complexity of the problem, generically, increases 
factorially with the size of the system. 
The problem of the quench dynamics in the integrable BCS model, 
(Eq.(\ref{pairing}), was thoroughly studied in~\cite{calabrese_quench}; 
see  also ~\cite{barankov-altshuler}). 
By employing the approach developed in~\cite{calabrese_dynamics} the authors 
proved, first, that the all the quench matrix is accessible by 
the Slavnov formula; then they proved that the quench dynamics occurs only 
along a relatively small subspace in the Hilbert space. 
Athough their results provide a hint that deviation from the 
mean-field regime emerges in the quench dynamics at finite size, 
further analysis seems to be required to unambiguously 
disclose the effect of mesoscopic pairing fluctuations.   

\acknowledgments 
We are pleased to thank A. Mastellone for discussions.





\begin{thebibliography}{100}
\bibitem{SKLYANIN-NOTATION} In the Ref.~\cite{SKLYANIN-CORRELATORS} 
	a different notation  was adopted  in that 
	the highest weight vector $s_j=1/2$  as vacuum was used.
\end{thebibliography}
\end{document}